\documentclass[aps,prd,nofootinbib]{revtex4}
\usepackage[colorlinks=true, pdfstartview=FitV, linkcolor=blue, citecolor=red, urlcolor=magenta]{hyperref}
\usepackage{graphicx}
\usepackage{latexsym}
\usepackage{amsmath}
\usepackage{amsfonts}
\usepackage{amssymb}
\usepackage{verbatim}
\usepackage{dsfont}
\usepackage{pbox}
\usepackage[dvipsnames, svgnames]{xcolor} 
\definecolor{darkblue}{RGB}{0, 0, 139} 

\newcommand{\be}{\begin{equation}}
\newcommand{\ee}{\end{equation}}
\newcommand{\bea}{\begin{eqnarray}}
\newcommand{\eea}{\end{eqnarray}}

\newcommand{\ben}{\begin{eqnarray}}
\newcommand{\een}{\end{eqnarray}}

\begin{document}

\title{Quantum Brownian motion induced by fluctuating boundaries and compactification}
\author{Eliza M. B. Guedes}
\email{ eliza.brito@academico.ufpb.br}
\author{Herondy Mota}
\email{hmota@fisica.ufpb.br}
\affiliation{Departamento de F\'isica, Universidade Federal da Para\'i­ba, Caixa Postal 5008, Jo\~ao Pessoa, Para\'iba, Brazil.}


\begin{abstract}
In this work, we investigate the quantum Brownian motion of a point charge arising as a consequence of two fluctuating point-like boundaries. The study considers Dirichlet, Neumann, and mixed boundary conditions imposed on a real massless scalar field. Additionally, we analyze the effects of a fluctuating compactification length on the random motion of the point charge, induced by the imposition of a quasi-periodic condition on the scalar field. By associating a wave function with the length scale of each system, we demonstrate that typical divergences, which commonly appear in scenarios with fixed boundaries and compactification size, are effectively smoothed out. This approach generalizes and extends previous results found in the literature, offering new insights into the regularization of divergences appearing in idealized systems.
\end{abstract}


\maketitle

\section{Introduction}
A point charge particle can undergo Brownian-like motion once it is coupled to a quantum field like the electric \cite{Yu:2004tu, Yu:2004gs, Ford:2005rs, Yu:2006tn, Bessa:2008pr, DeLorenci:2016jhd, DeLorenci:2019tyt, Bessa:2019aar, Lemos:2020ogj} or scalar field \cite{de2014quantum, camargo2018vacuum, camargo2019vacuum, Camargo:2020fxp, Ferreira:2023uxs}. The random trajectory described by the particle 
is induced by quantum vacuum fluctuations, resulting in nonzero velocity (VD) and position (PD) dispersions. This phenomenon, known as quantum Brownian motion (QBM), typically arises from the imposition of boundary conditions on the field, which modify its vacuum state.  The associated quantum vacuum fluctuations of a field, when altered,  
are responsible for a variety of effects \cite{milonni1994quantum, Ford:2005rs, BezerradeMello:2014phm, BezerradeMello:2011sm}, in addition to QBM.  The most well-known and experimentally verified of these is the Casimir effect \cite{bordag2009advances, Sparnaay:1958wg, Lamoreaux:1996wh, mohideen1998precision, Bressi:2002fr, PhysRevA.78.020101, PhysRevA.81.052115},  which strongly motivates the exploration of the consequences of quantum vacuum fluctuations in other contexts, such as those involving QBM.

The QBM induced by scalar vacuum fluctuations, as a consequence of Dirichlet, Neumann, and mixed boundary conditions in a system with two perfectly reflecting parallel planes, has been studied in Ref. \cite{Ferreira:2023uxs}. In the latter, the effects of a compactified dimension, introduced through a quasi-periodic condition that interpolates between the known periodic and anti-periodic ones via a phase parameter, are also investigated. Prior to the study conducted in Ref. \cite{Ferreira:2023uxs} on the scalar 
field case, other works considered a single plane with a Dirichlet boundary condition \cite{de2014quantum, camargo2018vacuum, camargo2019vacuum, Camargo:2020fxp}. These studies also highlighted additional aspects of the model that could be refined to achieve a more realistic representation of the system.

For instance, in Ref. \cite{camargo2019vacuum}, the authors investigate the violation of Huygens’ principle by considering a massive scalar field, as well as the effects of a smooth transition connecting distinct field states. To achieve this, a switch function with specific properties is utilized. For a massless scalar field, a similar approach is considered in Ref. \cite{camargo2018vacuum}, where two switching scenarios are adopted. Furthermore, in Ref. \cite{Camargo:2020fxp}, contributions from thermal fluctuations to the QBM in the massive case are analyzed, also considering switching functions. A common aspect of these investigations is that the introduction of switching functions makes it possible to regularize the analytic expressions describing velocity dispersion, which, in more idealized scenarios, present typical divergences. It is important to note that most of these works consider a massless scalar field model, as the technical aspects of the calculations become more tractable.

Regarding the smearing of the typical divergences appearing in idealized scenarios, where, for example, the boun\-daries are considered to be fixed and perfectly reflecting, another method besides the introduction of switching functions has been used in Ref. \cite{de2014quantum}. In this study, the authors consider a (1+1)-dimensional toy model for a massless scalar field subject to a Dirichlet boundary condition on an idealized point-like boundary. It is shown that the divergences arise both at the position of the boundary and for certain values of time, which are attributed to the time it takes for a light signal to complete a round trip between a given position and the boundary. This type of divergence is commonly referred to as flight-time divergence.

On the other hand, in Ref. \cite{de2014quantum}, the consequences of allowing the boundary's position to fluctuate about its mean value by adopting a Gaussian distribution are also discussed. In this scenario, the aforementioned divergences are regularized through a generalized hypergeometric function of the type $\;_{2}F_{2}\left[1,1;\frac{3}{2},2;-z\right] $. This analysis demonstrates that, when the boundary is treated as a quantum mechanical object with an associated wave function, the divergences tend to be eliminated. This idea has been previously explored in Ref. \cite{Ford:1998he}, which examines a system where vacuum energy density arises in (3+1) dimensions due to one- and two-plane boundaries. Consequently, the authors in Ref. \cite{Ford:1998he} showed that the existing divergences are also removed.

In the present work, we consider a toy model in (1+1)-dimensional Minkowski spacetime, where two perfectly reflecting point-like boundaries are taken into account. On these boundaries, we impose Dirichlet, Neumann, and mixed boundary conditions on a massless real scalar field coupled to a point charge undergoing QBM. We, thus, generalize the investigation conducted in Ref. \cite{de2014quantum} and show that the typical divergences appearing in all scenarios are removed by using a Gaussian distribution. Additionally, we consider a quasi-periodic condition as a way to compactify the spatial dimension, allowing the size of the compactification to fluctuate as well. In this case, it is also shown that the associated flight-time divergences are eliminated.

The structure of this paper is organized as follows. In Sec.\ref{Sec2}, we present the scalar field model and its coupling to the point charge. We obtain the scalar field Wightman functions for Dirichlet, Neumann, and mixed boundary conditions, as well as for the quasi-periodic condition. In Sec.\ref{secVD}, we calculate the VD for both fixed and fluctuating boundaries, and compactification size. We also obtain the PD in each case. In Sec.\ref{sec4}, we present our conclusions and final remarks. In this work, we use natural units such that $\hbar=c=1$.

\section{Scalar field Model, point charge dynamics and Wightman functions}\label{Sec2}
\subsection{Scalar field model and point charge dynamics}\label{sub2.1}
%
The system we want to consider is described by the total action, which includes the dynamics of a point charge coupled to a massless real scalar field. Since we are interested in analyzing the Brownian-like motion of this particle, we shall assume that it moves non-relativistically. Therefore, the non-covariant prescription for the action of the system is given by
\begin{equation}
S=\int dt \left(L_{\text{fld}} + L_{\text{par}} + L_{\text{int}}\right),
\label{actionT}
\end{equation}
where the Lagrangian for a massless scalar field $\phi(x,t)$ is characterized by 
\begin{equation}
L_{\text{fld}} =\int dx \left(\frac{1}{2}\partial_{\mu}\phi\partial^{\mu}\phi\right),
\label{la_fld}
\end{equation}
while, for a non-relativistic particle of mass $m$ in (1+1)-dimensional motion, the Lagrangian is written as
\begin{equation}
L_{\text{par}} = \frac{1}{2}mv^2,
\label{la_par}
\end{equation}
where $v=\frac{dx}{dt}$ is the velocity of the particle.

In addition, the interaction of the massless scalar field with a point particle of charge $g$ moving along a trajectory $\xi(t)$ is given by the following Lagrangian:
\begin{equation}
L_{\text{int}} = -g\int dx\;\phi\delta\left(x - \xi(t)\right).
\label{la_int}
\end{equation}

Now, upon varying the action \eqref{actionT} with respect to the scalar field, we obtain the equation of motion governing its dynamics as
\begin{equation}
\Box\phi(x,t) = - g\delta\left(x - \xi(t)\right),
\label{KG}
\end{equation}
where $\Box$ is the D'Alembertian operator in (1+1) dimensions defined as
\begin{equation}
\Box\equiv\left(\frac{\partial^2}{\partial t^2} - \frac{\partial^2}{\partial x^2}\right).
\label{DAO}
\end{equation}

On the other hand, the variation of the action \eqref{actionT} with respect to $\xi(t)$, the trajectory of the particle, provides the equation of motion for the non-relativistic point particle given by
\begin{equation}
m\frac{dv}{dt}=-g\frac{\partial\phi(x,t)}{\partial x}.
\label{EMP}
\end{equation}
As we can see the equation above takes into consideration only forces stemming from the presence of the scalar field, although we could also consider external forces from different natures \cite{Bessa:2008pr, ferreira2022quantum, JBFerreira:2023zwg}.

It is clear that Eq. \eqref{KG} is a non-homogeneous differential equation. As such, its general field solution can be expressed as the superposition of the corresponding homogeneous solution and a particular solution, the latter obtained using the retarded Green’s function associated with the massless scalar field \cite{fulling1989aspects}. The particular solution accounts for the back-reaction, which introduces dissipative effects into the particle's equation of motion \eqref{EMP}. This effect has been previously studied in Ref. \cite{gour1999will} within the context of QBM in (1+1)-dimensional free Minkowski spacetime. In our analysis, however, we will assume that the contribution of the particular solution to the general field is negligible, meaning dissipative effects are not considered. Therefore, the coupling between the massless scalar field and the massive point charge will be governed solely by the solution of Eq. \eqref{EMP} \cite{de2014quantum, Ferreira:2023uxs}, where $\phi$ represents the homogeneous solution of Eq. \eqref{KG}, to be discussed in the next subsection. The influence of boundary conditions on dissipative effects in the context of QBM will be explored in a separate study. Our primary focus here is to establish the conditions, at least in (1+1) dimensions, under which the typical divergences appearing in the VD of the particle can be smeared out. Notably, the emergence of these divergences is independent of dissipative effects.

As we wish to calculate the VD of the particle later on, the first step is to integrate Eq. \eqref{EMP} assuming that the particle starts from rest at $t=0$, that is, $v(t=0)=0$. This provides 
\begin{equation}
v =-\frac{g}{m}\int_0^{\tau}\partial_x\phi(x,t)\;dt.
\label{vel}
\end{equation}
Here, we have introduced the notation $\frac{\partial}{\partial x}\equiv\partial_x$. In general, the position coordinate in the argument of the scalar field in the expression above also depends on time. This makes the integration operation problematic, as, in principle, integration would only be feasible once the particle's trajectory is specified. To address this issue, we adopt the small displacement condition (SDC) hypothesis, which states that the particle's position does not vary significantly over time. Consequently, we can treat $x$ as a constant during the integration process in \eqref{vel} \cite{de2014quantum, Ferreira:2023uxs}.

We can still integrate Eq. \eqref{vel} to obtain an expression for the position of the particle. Upon assuming as initial condition $x(t=0)=0$, we find
\begin{equation}
 x = \int_0^{\tau} v(t_1)\;dt_1.
\label{pos}
\end{equation}

The previous two expressions will be paramount for the analysis performed in Sec.\ref{secVD}. Until then, let us proceed in the next subsection to calculate the Wightman function associated with the massless real scalar field considered here, under the influence of two point-like, perfectly reflecting boundaries. Additionally, for the same calculation, we will impose a quasi-periodic condition, effectively compactifying the \( x \)-direction and resulting in a \( S^1 \times \mathbb{R} \) topological spacetime.

%
\subsection{Field quantization and Wightman functions}
%
On the two boundaries, we wish to consider Dirichlet, Neumann, and mixed boundary conditions for the scalar field. Also, later in our analysis, we will introduce a quasi-periodic condition, which generalizes the well-known periodic and anti-periodic conditions. To begin with, we examine the solution to the homogeneous differential equation,
\begin{equation}
\Box\phi(x,t)\simeq0,
\label{homE}
\end{equation}
where we have neglected the particular solution, as previously stated. The solution to the above equation can be expressed as 
\begin{equation}
\phi_{\sigma}(x,t)=c_{\sigma}e^{-i\omega_{\sigma}t}\varphi_{\sigma}(x). 
\label{sol}
\end{equation}
Note that $\sigma$ stands for the modes of the field, $\omega_{\sigma}$ is the eigenfrequencies, $\varphi_{\sigma}(x)$ represents the spatial solution of the above expression and $c_{\sigma}$ is the normalization constant. The latter can be determined using the normalization condition in (1+1) dimensions, given by
\begin{equation}
2\omega_{\sigma}\int dx = \delta_{\sigma\sigma'},
\label{NC}
\end{equation}
where $\delta_{\sigma\sigma'}$ stands for either a Kronecker delta or a delta function depending on whether the mode is discrete or continuous, respectively. 

On two point-like boundaries separated by a distance $(b-a)$ and located at positions $x=a$ and $x=b$, we now impose Dirichlet (D), Neumann (N), and mixed boundary conditions on the scalar field. In Table \ref{t1}, we summarize these conditions along with their corresponding normalization constant $c_n$, discretized momentum $k_n$ and spatial solution $\varphi_n(x)$. 
It is clear that $\sigma=n$ for all three cases, as indicated. We also note that there are two possible configurations for mixed boundary conditions. Specifically, we can impose a Dirichlet condition on the scalar field at the boundary located at $x=a$ and a Neumann condition at the boundary located at $x=b$; we refer to this configuration as Dirichlet-Neumann (DN). Conversely, we can also have the Neumann-Dirichlet (ND) configuration. These cases are indicated in Table \ref{t1}.

We should point out that effects such as electrostatic interactions between the particle and the boundaries are neglected in our model. This is justified since the boundaries are treated as idealized and perfectly reflecting for the scalar field, and they do not induce classical potentials. Including such forces would require a more realistic model of the boundaries and is beyond the scope of the present analysis. However, over sufficiently long timescales, even weak classical forces could accumulate and modify the overall trajectory of the test particle, which may become relevant beyond the regime considered in this work.

The quantization rule requires us to construct the field operator in terms of a Fourier series of the type
\begin{eqnarray}
\hat{\phi}(w) = \sum_{\sigma}\left[a_{\sigma}\phi_{\sigma}(w) + a^{\dagger}_{\sigma}\phi^{*}_{\sigma}(w)\right],
\label{fieldOP}
\end{eqnarray}
where we have introduced the notation \( w = (x, t) \) for the spacetime coordinates, \( \phi_{\sigma}(w) \) are the mode solutions given by Eq. \eqref{sol}, and \( a_{\sigma} \) and \( a^{\dagger}_{\sigma} \) are the annihilation and creation operators, respectively, satisfying the standard commutation relation \( [a_{\sigma}, a^{\dagger}_{\sigma'}] = \delta_{\sigma\sigma'} \). Note that, although in our case the summation symbol in the above expression represents a sum over \( \sigma = n \) (as indicated in Table \ref{t1}), in general, it can also represent an integral when \( \sigma \) is continuous.

We can now proceed to calculate the positive frequency Wightman function, a quantity widely used in various contexts involving vacuum physics \cite{birrell1984quantum, fulling1989aspects}. To do so, we define the vacuum state \( |0\rangle \) of the scalar field such that \( a_{\sigma}|0\rangle = 0 \). Thus, the Wightman function is given by
\begin{eqnarray}
W(w,w')&=&\langle 0|\hat{\phi}(w)\hat{\phi}(w')|0\rangle\nonumber\\
&=&\sum_{\sigma}\phi_{\sigma}(w)\phi^{*}_{\sigma}(w')\nonumber\\
&=&\sum_{\sigma}|c_{\sigma}|^2e^{-i\omega_{\sigma}\Delta t}\varphi_{\sigma}(x)\varphi^{*}_{\sigma}(x'),
\label{WFD1}
\end{eqnarray}
where $\Delta t=t - t'$ and we have used Eqs. \eqref{fieldOP} and \eqref{sol}, along with the definition of vacuum state as previously described. 
\begin{table}[tbp]
\begin{tabular}{|l|l|l|l|l|}
\hline\hline
 & Boundary Condition & $\hspace{1.1cm}|c_{n}|^2$ & $\hspace{0.8cm}k_n$  & Solution $\varphi_n(x)$ \\ \hline
Dirichlet &\hspace{0.4cm}\pbox{20cm}{$\varphi(x=a)=0$\\ $\varphi(x=b)=0$} & \hspace{0.7cm}$\frac{1}{\omega_{n}(b-a)}$ & \pbox{20cm}{$k_n=\frac{n\pi}{(b-a)}$\\ $n= 1, 2, 3, ...$} & \hspace{0.12cm}$\sin[k_n(x-a)]$ \\ \hline
Neumann & \hspace{0.3cm}\pbox{20cm}{$\partial_x\varphi(x=a)=0$\\ $\partial_x\varphi(x=b)=0$} & \pbox{20cm}{$\frac{1}{2\omega_{n}(b-a)}$ for $n=0$ \\ $\frac{1}{\omega_{n}(b-a)}$ \;\;for $n\geq 1$} & \pbox{20cm}{$k_n=\frac{n\pi}{(b-a)}$\\ $n= 0, 1, 2, 3, ...$} & \hspace{0.12cm}$\cos[k_n(x-a)]$ \\ \hline
Mixed & \hspace{0.3cm}\pbox{20cm}{$\varphi(x=a)=0$\\ $\partial_x \varphi(x=b)=0$ \\ or \\ $\varphi(x=b)=0$\\ $\partial_x \varphi(x=a)=0$} & \hspace{0.7cm}$\frac{1}{\omega_{n}(b-a)}$ & \pbox{20cm}{$k_n=\frac{(2n+1)\pi}{2(b-a)}$\\ $n= 0, 1, 2, 3, ...$} & \hspace{0.12cm}\pbox{20cm}{$\sin[k_n(x-a)]$\\ or \\$\cos[k_n(x-a)]$}\\ \hline\hline
\end{tabular}%
\caption{Dirichlet, Neumann and mixed boundary conditions, along with their corresponding normalization constant $c_n$, discretized momentum $k_n$ and spatial solution $\varphi_n(x)$. Note that for all boundary conditions considered, the eigenfrequencies satisfy the dispersion relation \(\omega_n = |k_n|\), since we are dealing with a real massless scalar field in flat spacetime.}
\label{t1}
\end{table}

For Dirichlet, Neumann and mixed boundary conditions, we can make use of Table \ref{t1} to calculate the Wightman function \eqref{WFD1}. It turns out that with the help of the delta function property $\int\delta(z-z_0)g(z)dz=g(z_0)$ and also of the identity \cite{arfken2005mathematical}
\begin{equation}
\sum_{n=-\infty}^{\infty}e^{idn}=2\pi\sum_{n=-\infty}^{\infty}\delta(d - 2\pi n),
\label{identity1}
\end{equation}
the Wightman function for the three boundary condition cases can be put in the form \cite{Ford:1998he}
\begin{eqnarray}
W(w,w')=\frac{1}{4\pi}\sum_{n=-\infty}^{\infty}\int_{-\infty}^{\infty} d\kappa\left[\delta_n^{(\text{i})}\frac{e^{-i\omega_{\kappa}\Delta t + i\kappa(\Delta x + 4bn)}}{\omega_{\kappa}} + \gamma_n^{(\text{i})}\frac{e^{-i\omega_{\kappa}\Delta t + i\kappa(\Delta\bar{x} + 2b(2n+1))}}{\omega_{\kappa}}\right],
\label{WFComp}
\end{eqnarray}
where $\Delta x=x - x'$ and  $\Delta\bar{x}=x + x'$. Here, \(\kappa\) represents the continuous version of the discretized momenta \(k_n\), appropriate when transitioning from mode sums to integrals in the continuum limit.
The coefficients $(\delta_n^{(\text{i})}, \gamma_n^{(\text{i})})$ for each boundary condition case are defined as
\begin{eqnarray}
\delta_n^{(\text{i})} &=& \left[\delta_n^{(\text{D})}, \delta_n^{(\text{N})}, \delta_n^{(\text{DN})}, \delta_n^{(\text{ND})}\right]\nonumber\\
&=&\left[+1, +1, (-1)^n, (-1)^n\right],
\label{coef1}
\end{eqnarray}
and also
\begin{eqnarray}
\gamma_n^{(\text{i})} &=& \left[\gamma_n^{(\text{D})}, \gamma_n^{(\text{N})}, \gamma_n^{(\text{DN})}, \gamma_n^{(\text{ND})}\right]\nonumber\\
&=&\left[-1, +1, (-1)^{n+1}, (-1)^n\right].
\label{coef2}
\end{eqnarray}

Note that, for the sake of simplicity, in Eq. \eqref{WFComp}, we have set $a=-b$, making the boundaries symmetrically located about $x=0$. In principle, we could consider the more general case where the boundaries are located at arbitrary positions 
$a$ and $b$. However, this would complicate the mathematical analysis of boundary fluctuations and obscure the mechanism by which the existing divergences can be smoothed out. Therefore, by assuming $a=-b$, the analysis becomes more straightforward.

An additional observation to note is that we have introduced a cutoff parameter $M$ in transitioning from Eq. \eqref{WFD1} to Eq. \eqref{WFComp}. This allows the eigenfrequencies to be expressed as $\omega_{\kappa}=\sqrt{\kappa^2 + M^2}$. This parameter is necessary to circumvent the infrared divergence that arises when attempting to solve the integral in Eq. \eqref{WFComp}. Such divergence is characteristic of massless scalar field theories in (1+1)-dimensional spacetime \cite{birrell1984quantum}. However, as we will see, this cutoff parameter does not affect the calculation of the VD of the point particle. This is because the VD involves taking spatial derivatives of the Wightman function, which ultimately eliminates the dependence on 
$M$. In a massive scalar field theory, this cutoff parameter corresponds to the mass of the field.

Let us now turn to the problem of solving the integral in Eq.~\eqref{WFComp}. This is possible, for instance, by making use of the integral identity
\begin{eqnarray}
\frac{e^{-\omega\Delta T}}{\omega}=\frac{2}{\sqrt{\pi}}\int_0^{\infty}dse^{-\omega^2s^2-\frac{\Delta T^2}{4s^2}},
\label{identity2}
\end{eqnarray}
where in our case $\Delta T=i\Delta t$ (Wick rotation) and $\omega=\omega_{\kappa}$. Note that although the integral representation above, obtained after the Wick rotation, \( \Delta T = it \), appears formally divergent for real \( t \), it is to be interpreted via analytic continuation in the complex plane. This continuation allows the transition to Euclidean space, where path integrals and Green's functions are often better defined. Such procedures are standard in quantum field theory, where time-ordered Green's functions are recovered from their Euclidean counterparts through inverse Wick rotation, after the integral is performed.

In terms of the cutoff parameter $M$, Eq.~\eqref{identity2} leads to the renormalized Wightman function
\begin{eqnarray}
W_{\text{R}}(w,w')=&&\frac{1}{2\pi}\sideset{}{'}\sum_{n=-\infty}^{\infty}\delta_n^{(\text{i})}K_0\left(M\sqrt{\left(\Delta x + 4bn\right)^2 - \Delta t^2}\right) \nonumber\\
&+& \frac{1}{2\pi}\sum_{n=-\infty}^{\infty}\gamma_n^{(\text{i})}K_0\left(M\sqrt{\left(\Delta\bar{x} + 2b(2n+1)\right)^2 - \Delta t^2}\right), 
\label{WFCompF}
\end{eqnarray}
where \( K_{\mu}(z) \) is the Macdonald function, also known as the modified Bessel function of the second kind\cite{abramowitz1970handbook}. The prime notation in the sum on the first line of the above expression indicates that the term \( n = 0 \) has been excluded. This term corresponds to the Minkowski contribution, which diverges in the coincidence limit \( w \rightarrow w' \). This limit is to be taken after completing the necessary operations. For completeness, we write the Minkowski contribution as follows:
\begin{equation}
W_{\text{M}}(w,w')=\frac{1}{2\pi}K_0\left(M\sqrt{\Delta x^2 - \Delta t^2}\right).
\label{WFM}
\end{equation}
Since the asymptotic limit for small arguments of the Bessel function above goes as $K_{0}(z)\simeq -\ln(z)$ \cite{abramowitz1970handbook}, the Wightman function generically assumes the form
\begin{equation}
W(w,w')\simeq - \frac{1}{2\pi}\ln\left(M\Delta r\right). 
\label{logdiv}
\end{equation}
This shows that the Minkowski contribution logarithmically diverges in the coincidence limit $\Delta r\rightarrow 0$. In contrast, for any other $n$ contribution in Eq. \eqref{WFCompF} this does not happen.

Another important contribution to be noted is the one corresponding to a single boundary, which arises from the term \( n = 0 \) in the second line of Eq.  \eqref{WFCompF}. This can be expressed as
\begin{equation}
W_{\text{1b}}(w,w')=\frac{1}{2\pi}K_0\left(M\sqrt{(\Delta\bar{x} + 2b)^2 - \Delta t^2}\right).
\label{WF1p}
\end{equation}
Although this expression does not diverge in the coincidence limit \( w \rightarrow w' \), it still assumes the asymptotic form in Eq. \eqref{logdiv}, since we are interested in the massless scalar field case, obtained in the limit \( M \rightarrow 0 \). 
When adopting Dirichlet boundary conditions, the VD of the particle in the context of a single perfectly reflecting boundary has been studied in Ref. \cite{de2014quantum}, where the authors also discuss the effects of boundary fluctuations. Thus, our results aim to generalize the analysis of Ref. \cite{de2014quantum} by also considering Neumann and mixed boundary conditions.

To conclude this section, we will also examine the case where the scalar field satisfies a quasi-periodic condition, given by
\begin{eqnarray}
\varphi(x)=e^{-2\pi i\beta}\varphi(x+L),
\label{quasi_condition}
\end{eqnarray}
where \( L \) is a length parameter and \( 0 \leq \beta < 1 \). The latter corresponds to the periodic condition when \( \beta = 0 \) and the anti-periodic condition when \( \beta = 1/2 \). The other values within the aforementioned interval represent interpolations between these two cases \cite{Barone:2003nk}.

After imposing the condition above on the scalar field solution in Eq. \eqref{sol}, we obtain
\begin{eqnarray}
\varphi(x) = e^{ ik_n x},
\label{quasi_condition}
\end{eqnarray}
for the spatial section of the scalar field $\phi(w)$. Note that the discretized momentum in this case is written as
\begin{equation}
k_n=\frac{2\pi}{L}(n+\beta),
\label{quasi_momentum}
\end{equation}
where $n=0,\pm1, \pm 2, \pm 3,...\;$. Moreover, by making use of the normalization condition \eqref{NC}, we obtain that the normalization constant is given by 
\begin{equation}
c_n=\frac{1}{\sqrt{2\omega_n L}}.
\label{NCQP}
\end{equation}

The two-point Wightman function can now be calculated using Eqs. \eqref{WFD1}, \eqref{quasi_condition}, and \eqref{NCQP}. Additionally, we can apply the delta function property \( \int \delta(z - z_0) g(z) \, dz = g(z_0) \) along with the identity in Eq. \eqref{identity1} to express the Wightman function in the form \cite{Ford:1998he}
\begin{eqnarray}
W(w,w')=\frac{1}{4\pi}\sum_{n=-\infty}^{\infty}e^{-2\pi i\beta n} \int_{-\infty}^{\infty}d\kappa\frac{e^{-i\omega_{\kappa}\Delta t + i\kappa(\Delta x + Ln)}}{\omega_{\kappa}}. 
\label{WFQuasi}
\end{eqnarray}
Then, with the help of the integral identity \eqref{identity2} and also of the cutoff parameter $M$ present in the eigenfrequencies $\omega_{\kappa}$, we obtain the renormalized Wightman function in the quasi-periodic case as
\begin{eqnarray}
W_{\text{R}}(w,w')=\frac{1}{2\pi}\sideset{}{'}\sum_{n=-\infty}^{\infty}e^{-2\pi i\beta n}K_0\left(M\sqrt{\left(\Delta x + Ln\right)^2 - \Delta t^2}\right), 
\label{WFQuasi2}
\end{eqnarray}
where we have subtracted the divergent Minkowski contribution shown in Eq. \eqref{WFM}. Hence, the prime notation above indicates the absence of the term $n=0$ in the sum, referring to this divergence. The expression above is finite in the coincidence limit $w\rightarrow w'$, but logarithmically diverges in the 
limit $M\rightarrow 0$, according to the asymptotic form in Eq. \eqref{logdiv}.

So far, we have obtained the expression \eqref{WFCompF} for the renormalized Wightman function associated with Dirichlet, Neumann, and mixed boundary conditions, as well as Eq. \eqref{WFQuasi2} for the renormalized Wightman function associated with the quasi-periodic condition. These expressions, in the limit \( M \rightarrow 0 \), assume the asymptotic form given by Eq. \eqref{logdiv}, which diverges exactly at \( M = 0 \). As a result, they represent cutoff-dependent expressions for the massless scalar field. However, to calculate the VD of the particle, we will need to take spatial derivatives of the Wightman functions, which eliminates the cutoff dependence. This will be demonstrated in the next section for the calculation of the PD of the particle.

The VD of the particle in (3+1)-dimensional Minkowski spacetime, under the conditions adopted here, has been studied in Ref. \cite{Ferreira:2023uxs}. There, it is shown that the Wightman functions do not exhibit an infrared divergence problem and, as a result, yield finite expressions for the massless scalar field cases after the Minkowski contribution is removed. Ref. \cite{Bessa:2019aar} suggests that the effect of the boundary conditions listed in Table \ref{t1} in problems involving QBM serves as an indicator of the global inhomogeneity of space (see also \cite{Lemos:2020ogj, Lemos:2021jzy} for similar discussions). Thus, in (3+1) dimensions, the presence of two parallel planes breaks the homogeneity and isotropy of the spatial section of spacetime. We may infer that a similar effect occurs in (1+1) dimensions. On the other hand, the quasi-periodic condition provides a method for compactifying a specific direction, such as the \( x \)-direction assumed here. As a result, in (3+1) dimensions, the topology is modified to \( \mathbb{R}^3 \times S^1 \). When only (1+1) dimensions are considered, the spacetime topology becomes \( \mathbb{R}^1 \times S^1 \), which is the case in our analysis. All four conditions discussed here are of significant interest and have practical applications. For instance, Ref. \cite{alves2000spontaneous} employs mixed boundary conditions to study the spontaneous emission of a two-level system. In contrast, the quasi-periodic condition can be used to simulate an interaction in the system analogous to the well-known Aharonov-Bohm effect \cite{de2012topological, kretzschmar1965must}.

\section{Velocity and position dispersions}
\label{secVD}
%
We are interested in analyzing the QBM induced by quantum vacuum fluctuations of the scalar field. To do so, we must consider the impact of promoting the field to an operator, as described in Eq. \eqref{fieldOP}, on the velocity of the point particle given in Eq. \eqref{vel} and on its position in Eq. \eqref{pos}. As a result, these quantities are quantized in such a way that the vacuum expectation value (VEV) of the velocity operator vanishes, i.e., \( \langle 0|\hat{v}|0\rangle = 0 \), and consequently, \( \langle 0|\hat{x}|0\rangle = 0 \). Nevertheless, there exist nonzero induced dispersions of these quantities, which, in this case, are equivalent to the VEV of the squared particle velocity and position \cite{Ferreira:2023uxs, de2014quantum}. In terms of the renormalized Wightman function, the VD of the particle is then given by
\begin{equation}
 \langle(\Delta v)^2\rangle_{\text{R}}=\frac{g^2}{m^2}\int_0^{\tau}dt\int_0^{\tau}dt' \partial_x\partial_{x'}W_{\text{R}}(w,w'),
\label{D_vel}
\end{equation}
where $\tau$ is an arbitrary time and we have introduced the notation $\langle 0|(...)|0\rangle = \langle(...)\rangle$.

Equivalently, in terms of the renormalized Wightman function, the PD of the particle is written as 
\begin{equation}
 \langle(\Delta x)^2\rangle_{\text{R}}=\frac{g^2}{m^2}\int_0^{\tau}dt_2\int_0^{\tau}dt_1\int_0^{t_2}dt'\int_0^{t_1}dt\partial_x\partial_{x'}W_{\text{R}}(w,w').
\label{D_pos}
\end{equation}

Using the expressions above, we can apply the results from the previous section, where we calculated the renormalized Wightman function for Dirichlet, Neumann, and mixed boundary conditions, as well as the renormalized Wightman function for the quasi-periodic condition. We will explore this in the following subsections.

\subsection{Dirichlet, Neumann and mixed boundary conditions}
%
For the calculation of the VD of the particle, the final closed-form expression of the Wightman function is typically used. In our case, the expressions to be considered are those in Eqs. \eqref{WFCompF} and \eqref{WFQuasi2}, along with their corresponding asymptotic form \eqref{logdiv}. For the massless scalar field, the operations of differentiation (where the cutoff \( M \) disappears) and integration are performed in Eq. \eqref{D_vel} \cite{Ferreira:2023uxs, de2014quantum}. However, we will adopt a different approach to obtain the VD, which proves to be more convenient for analyzing the fluctuations of the boundaries, as demonstrated below.

We will first consider the Wightman function for Dirichlet, Neumann, and mixed boundary conditions in the form of Eq. \eqref{WFComp}. After performing the integrations and derivatives required in Eq. \eqref{D_vel} and taking the coincidence limit \( w' \rightarrow w \), we obtain
\begin{eqnarray}
\langle(\Delta v)^2\rangle_{\text{R}}=\frac{g^2}{2\pi m^2} \int_{-\infty}^{\infty}d\kappa\frac{\kappa^2\left[1-\cos(|\kappa|\tau)\right]}{|\kappa|^3}\left[\sideset{}{'}\sum_{n=-\infty}^{\infty}\delta_n^{(\text{i})}e^{4i\kappa bn} - \sum_{n=-\infty}^{\infty}\gamma_n^{(\text{i})}e^{2 i\kappa (x + b(2n+1))}\right],
\label{VDR1}
\end{eqnarray}
where we have used $\omega_{\kappa}=|\kappa|$. Note that the cutoff \( M \) is not necessary to proceed with the calculation, so we have set \( M = 0 \). Thus, the integral above can be evaluated using the relation given by
\begin{eqnarray}
I &=& \int_{0}^{\infty}d\kappa\frac{\left[1-\cosh(\kappa T)\right]\cos(\kappa z)}{\kappa}\nonumber\\
&=& -\frac{1}{2}\ln\left(\frac{ z^2}{ z^2 + T^2}\right).
\label{int1}
\end{eqnarray}
In our case, $T=i\tau$ and $z=4bn$ for the first and $ z=2(x + b(2n+1))$ for the second terms on the r.h.s. of Eq. \eqref{VDR1}, respectively.  With this, the renormalized VD of the particle for fixed boundaries is found to be
\begin{eqnarray}
\langle(\Delta v)^2\rangle_{\text{R}}=&-&\frac{g^2}{2\pi m^2}\sum_{n=1}^{\infty}\delta_n^{(\text{i})}\ln\left[\left(\frac{16n^2b^2}{16n^2b^2 - \tau^2}\right)^2\right]\nonumber\\
&-&\frac{g^2}{4\pi m^2}\sum_{n=-\infty}^{\infty}\gamma_n^{(\text{i})}\ln\left[\left(\frac{[4nb + 2(x+b)]^2}{[4nb + 2(x+b)]^2 - \tau^2}\right)^2\right],
\label{VDR2}
\end{eqnarray}
where the term \( n = 0 \) in the second line corresponds to the single-boundary contribution, as investigated in Ref. \cite{de2014quantum}. The result above, therefore, generalizes the findings presented in \cite{de2014quantum}, which were based solely on Dirichlet boundary condition. An intriguing aspect of this expression is that the VD can assume negative values (see Fig.\ref{figure1}). Investigations into this phenomenon suggest that negative dispersion arises from the reduction in quantum uncertainty induced by the presence of boundaries \cite{Yu:2004tu, Yu:2004gs}. This is known in the literature as subvacuum effect \cite{camargo2019vacuum, Camargo:2020fxp, de2014quantum}.

Some important remarks are in order regarding the above expression for the VD. Upon closer inspection, we can observe divergences associated with certain values of \( \tau \) and \( x \). For instance, in the first line of Eq. \eqref{VDR2}, there are typical flight-time divergences at \( \tau = 4bn \). Given that the distance between the boundaries is \( 2b \), this type of divergence is attributed to the time a light signal takes to complete a round trip between the boundaries \cite{de2014quantum, Ferreira:2023uxs}. Additionally, there are position-dependent flight-time divergences at \( \tau = 4nb + 2(x + b) \), arising from the term in the second line. In this case, the round trip begins at a specific position between the boundaries. On the other hand, in the numerator of the \( \ln(z) \) term in the second line, divergences occur precisely at the positions of the boundaries, \( x = -b \) and \( x = b \), when \( n = 0 \) and \( n = -1 \), respectively.
\begin{figure}[h]
\includegraphics[scale=0.3]{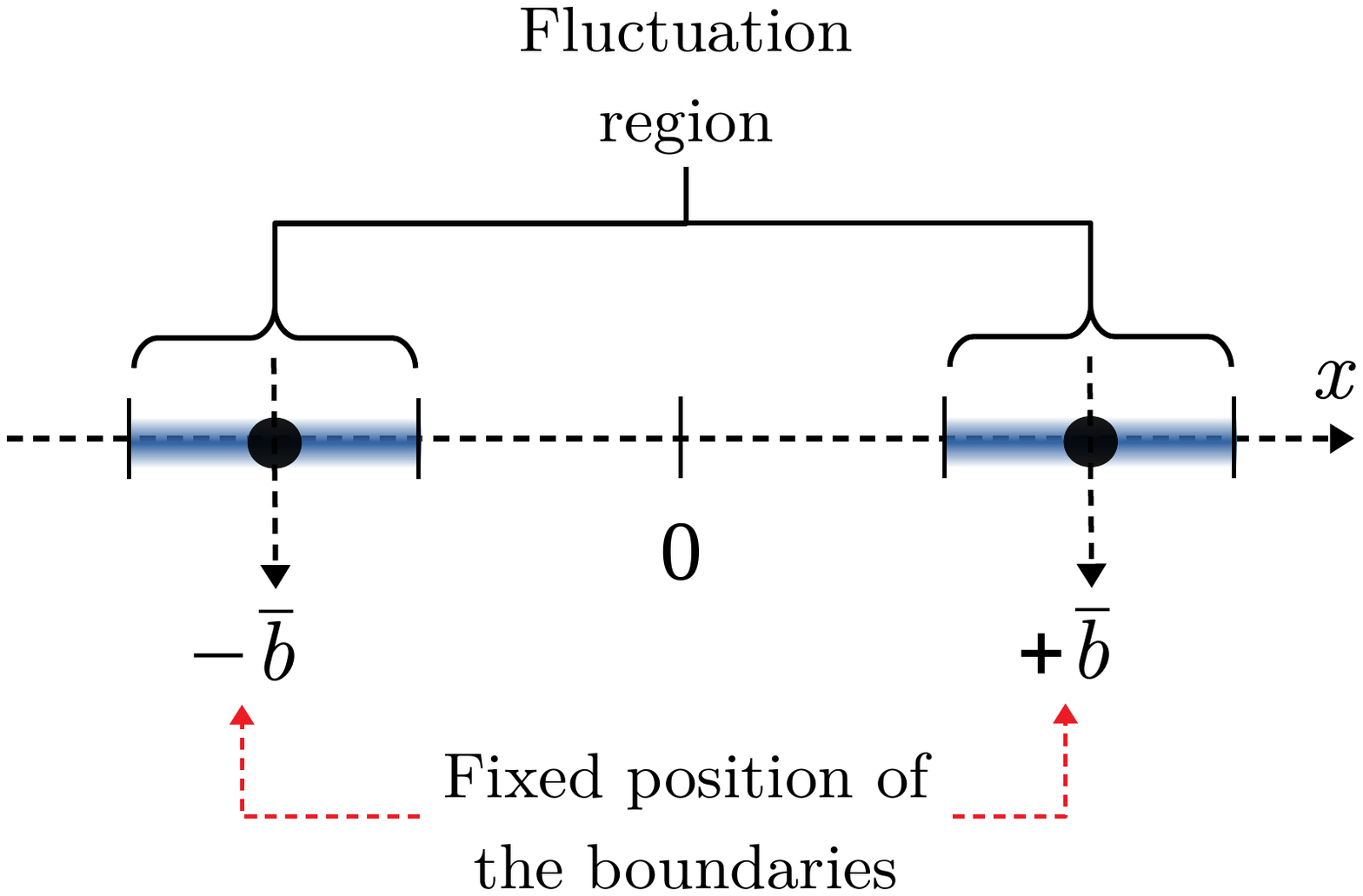}
\caption{Illustrative view of the fluctuating boundaries about their fixed mean positions \( -\bar{b} \) and \( \bar{b} \).}
\label{figure0}
\end{figure}

It is argued that the divergences mentioned above are related to the idealization of the boundaries. In principle, one way to smear out these divergences is to associate a wave function with the position of the boundaries, allowing them to fluctuate \cite{Ford:1998he, de2014quantum}. Let us then consider that the position \( b \) is a random variable and can fluctuate by a quantity \( \epsilon \) about its mean value \( \bar{b} \) (see Fig.\ref{figure0}), that is,
\begin{eqnarray}
b=\bar{b}(1 + \epsilon),
\label{b_fluctuation}
\end{eqnarray}
where the normalized probability distribution adopted here is gaussian, which can be written as
\begin{eqnarray}
g(\epsilon)=\frac{1}{\sqrt{2\pi}\sigma}e^{-\frac{\epsilon^2}{2\sigma^2}},
\label{PDG}
\end{eqnarray}
where $\sigma$ is a constant related to the width of the gaussian. Thus, the mean value of a function $\mathcal{M}(\epsilon)$ is given by
\begin{eqnarray}
\overline{\mathcal{M}}=\frac{1}{\sqrt{2\pi}\sigma}\int_{-\infty}^{\infty}\mathcal{M}(\epsilon)e^{-\frac{\epsilon^2}{2\sigma^2}}d\epsilon.
\label{MV}
\end{eqnarray}
Using the probability distribution in Eq. \eqref{PDG} along with Eq. \eqref{MV}, it is straightforward to verify that the mean value of \( \epsilon \) vanishes, i.e., \( \bar{\epsilon} = 0 \), and that its uncertainty is given by \( \overline{\Delta\epsilon^2} = \sigma^2 \). Additionally, it can be shown that, by using Eqs. \eqref{b_fluctuation} and \eqref{MV}, \( \bar{b} \) is indeed the mean value of \( b \).
\begin{figure}[h]
\includegraphics[scale=0.55]{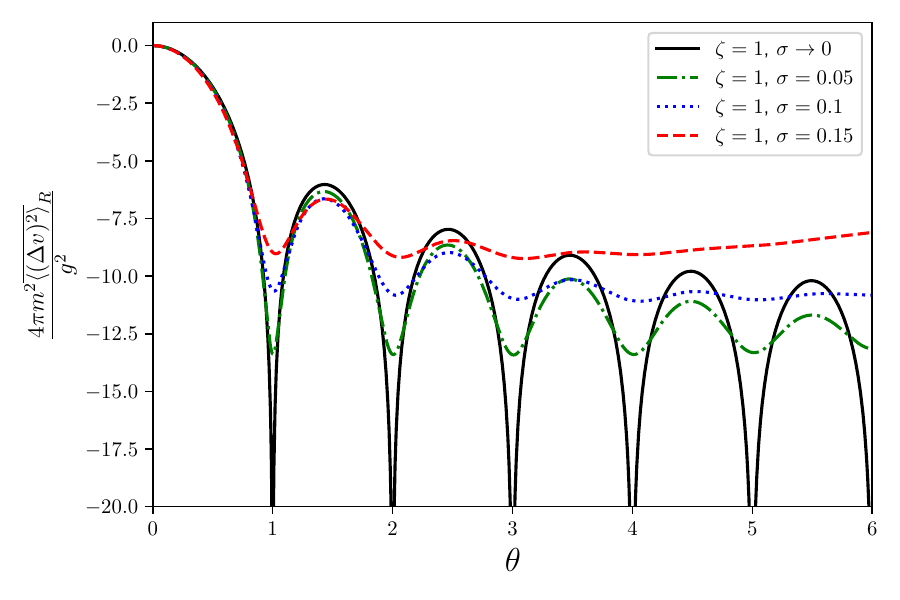}
\includegraphics[scale=0.55]{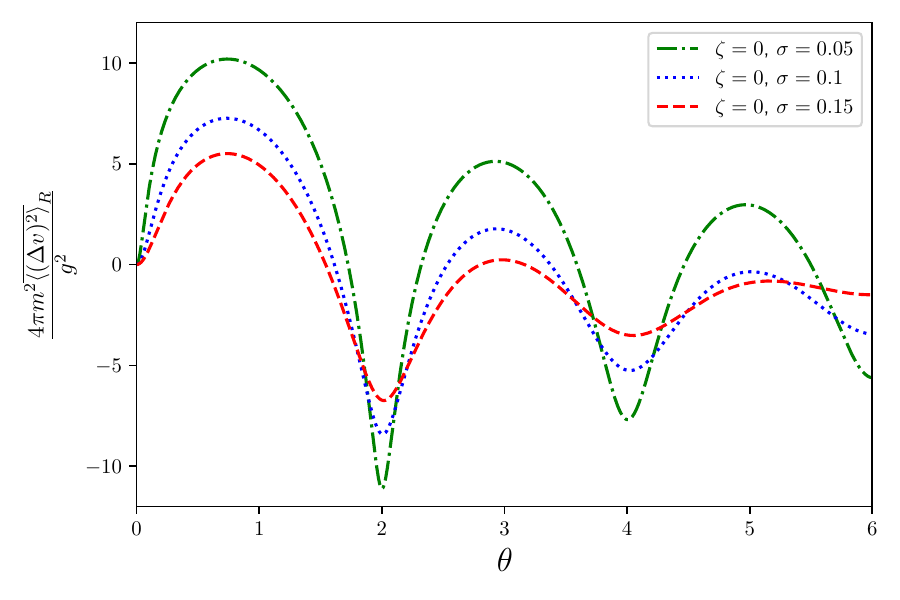}
\includegraphics[scale=0.55]{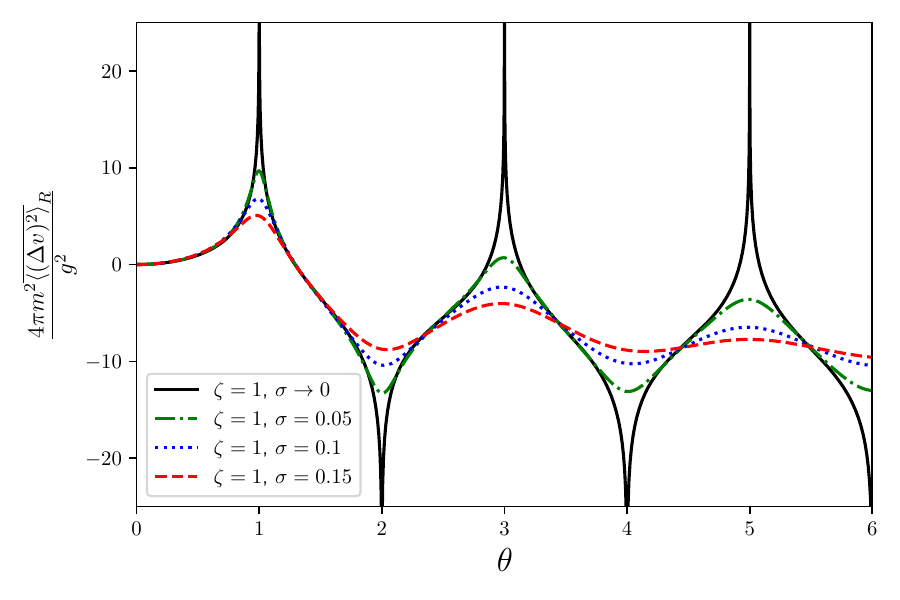}
\includegraphics[scale=0.55]{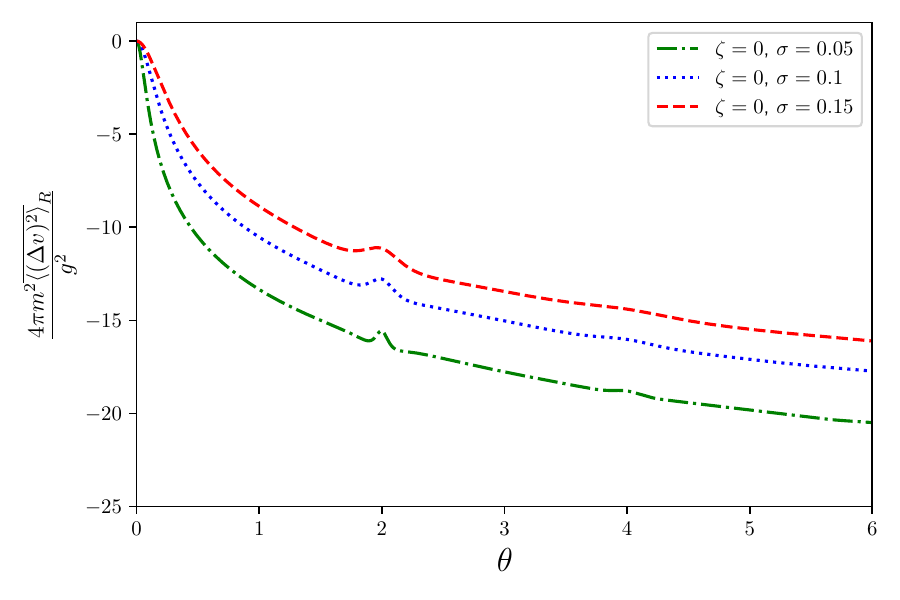}
\caption{Plots of the VD \eqref{VDRFluc} as a function of \( \theta \), considering only Dirichlet (top) and Neumann (bottom) boundary conditions. It is evident that, for finite values of \( \sigma \), the flight-time divergences and the divergences at the boundaries are smoothed out.}
\label{figure1}
\end{figure}

We can now introduce fluctuations in the position \( b \) of the boundaries by considering the VD in the form of Eq. \eqref{VDR1} and using Eq. \eqref{b_fluctuation}. From Eq. \eqref{MV}, we obtain the mean value of the exponential terms in Eq. \eqref{VDR1} in the general form \( \overline{e^{i\kappa p\epsilon}} = e^{-\frac{\kappa^2 p^2\sigma^2}{2}} \). Subsequently, in Eq. \eqref{VDR1}, we are left with the integral in \( \kappa \), which can be evaluated using the result below:
\begin{align*}
I&=\sum_{j=+,-} \int_{0}^{\infty}d\kappa\frac{\left[1-\cosh(\kappa T)\right]e^{iyj\kappa - z^2\kappa^2}}{\kappa}\nonumber\\
&=
\begin{aligned}[t]
&-\frac{y^2}{2z^2}\;_{2}F_{2}\left[1,1;\frac{3}{2},2;-\frac{y^2}{4z^2}\right]\nonumber\\
&- \frac{(T - iy)^2}{4z^2}\;_{2}F_{2}\left[1,1;\frac{3}{2},2;\frac{(T - iy)^2}{4z^2}\right]\nonumber\\
&- \frac{(T + iy)^2}{4z^2}\;_{2}F_{2}\left[1,1;\frac{3}{2},2;\frac{(T + iy)^2}{4z^2}\right],
\end{aligned}
\label{int2}
\end{align*}
where the sum over \( j \) has been introduced to account for the fact that the integral in \( \kappa \) in Eq. \eqref{VDR1} is divided into two contributions. After adjusting the parameters \( T \), \( y \), and \( z \) for our case and applying the change of variables,
\begin{eqnarray}
\theta = \frac{\tau}{2\bar{b}},\qquad\qquad\qquad\zeta=\frac{x}{\bar{b}}+1,
\label{changeV}
\end{eqnarray}
we find the VD of the particle as a consequence of the fluctuating boundaries as
\begin{eqnarray}
\overline{\langle(\Delta v)^2\rangle}_{\text{R}}=-\frac{g^2}{4\pi m^2\sigma^2}\left[\sideset{}{'}\sum_{n=-\infty}^{\infty}\delta_n^{(\text{i})}P_n(\theta) -\sum_{n=-\infty}^{\infty}\gamma_n^{(\text{i})}\frac{S_n(\theta,\zeta)}{(2n + 1)^2}\right],
\label{VDRFluc}
\end{eqnarray}
where for the first term on the r.h.s. we have defined 
\begin{eqnarray}
P_n(\theta)=&-&\left(1 + \frac{\theta}{2n}\right)^2\;_{2}F_{2}\left[1,1;\frac{3}{2},2;-\frac{\left(1 + \frac{\theta}{2n}\right)^2}{2\sigma^2}\right]\nonumber\\
&-&\left(1 - \frac{\theta}{2n}\right)^2\;_{2}F_{2}\left[1,1;\frac{3}{2},2;-\frac{\left(1 - \frac{\theta}{2n}\right)^2}{2\sigma^2}\right]\nonumber\\
&+&2\;_{2}F_{2}\left[1,1;\frac{3}{2},2;-\frac{1}{2\sigma^2}\right],
\label{p1}
\end{eqnarray}
while for the second term we have
\begin{eqnarray}
S_n(\theta,\zeta)=&-&(\zeta + 2n + \theta)^2\;_{2}F_{2}\left[1,1;\frac{3}{2},2;-\frac{(\zeta + 2n + \theta)^2}{2(2n + 1)^2\sigma^2}\right]\nonumber\\
&-&(\zeta + 2n - \theta)^2\;_{2}F_{2}\left[1,1;\frac{3}{2},2;-\frac{(\zeta + 2n - \theta)^2}{2(2n + 1)^2\sigma^2}\right]\nonumber\\
&+&2(\zeta + 2n)^2\;_{2}F_{2}\left[1,1;\frac{3}{2},2;-\frac{(\zeta + 2n)^2}{2(2n + 1)^2\sigma^2}\right],
\label{p2}
\end{eqnarray}
where both terms are given in terms of generalized hypergeometric functions of the type $\;_{2}F_{2}[1,1;\frac{3}{2},2; -z]$. Note that the divergences appearing in Eq. \eqref{VDR2} take the form \( \theta = 2n \) and \( \theta = 2n + \zeta \) in the new variables. Additionally, we should recall that the expression in Eq. \eqref{VDRFluc} is a compact representation encompassing Dirichlet, Neumann, and mixed boundary conditions. The specific case to be considered will depend on the parameters \( \delta_n^{(\text{i})} \) and \( \gamma_n^{(\text{i})} \), defined in Eqs. \eqref{coef1} and \eqref{coef2}, respectively. Regarding the structure of the VD in Eq. \eqref{VDRFluc}, we observe that the term \( S_0(\theta, \zeta) \) corresponds to the single-boundary contribution discussed in Ref. \cite{de2014quantum}. 
\begin{figure}[h]
\includegraphics[scale=0.55]{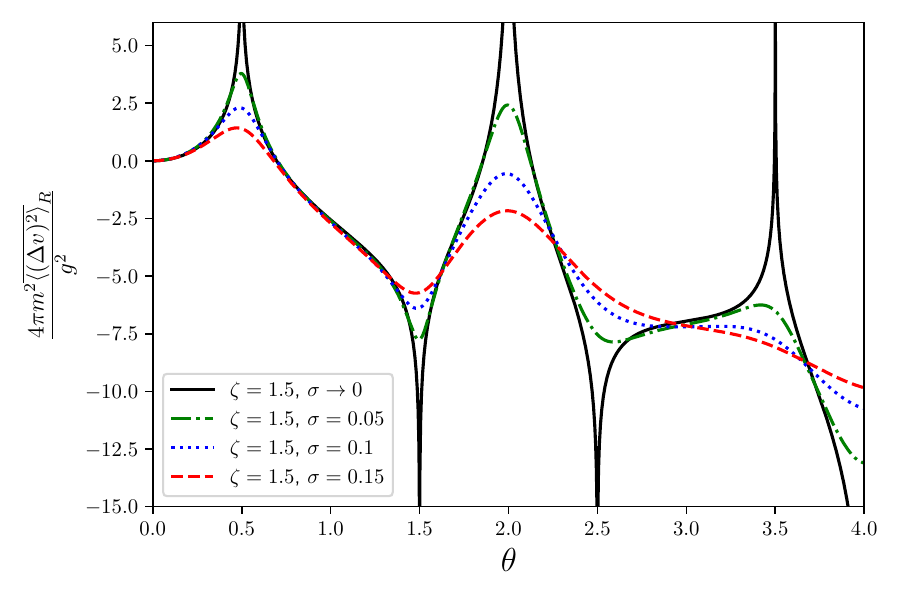}
\includegraphics[scale=0.55]{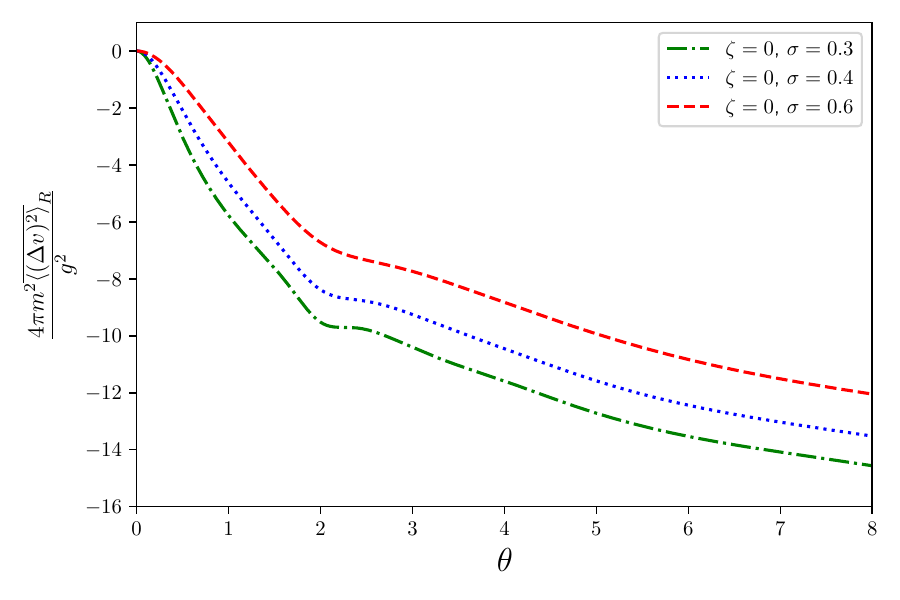}
\includegraphics[scale=0.55]{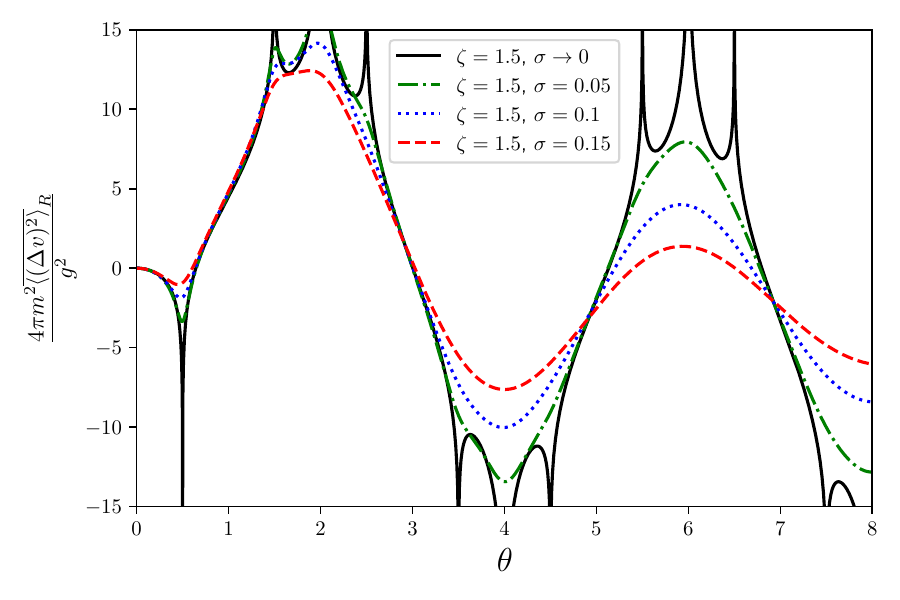}
\includegraphics[scale=0.55]{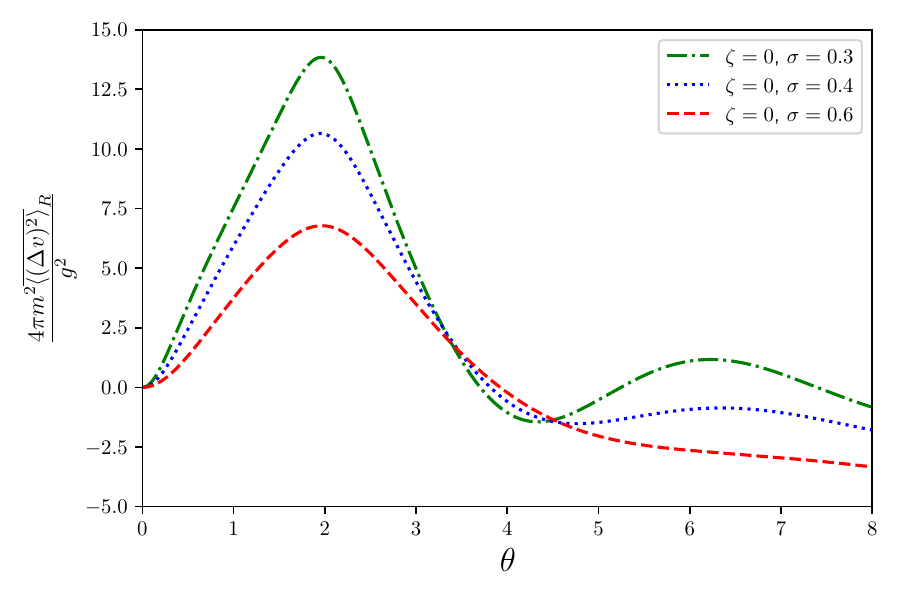}
\caption{Plots of the VD \eqref{VDRFluc} as a function of \( \theta \), considering only DN (top) and ND (bottom) conditions. It is evident that, for finite values of \( \sigma \), the flight-time divergences and the divergences at the boundaries are smoothed out.}
\label{figure2}
\end{figure}

The expression in Eq. \eqref{VDRFluc} is now finite at the boundaries, i.e., at \( \zeta = 0 \) (\( x = -\bar{b} \)) and \( \zeta = 2 \) (\( x = \bar{b} \)). With respect to \( \theta \), this is illustrated in Fig.\ref{figure1} for Dirichlet (top right) and Neumann (bottom right) boundary conditions, considering \( \zeta = 0 \) and different values of \( \sigma \). The case \( \zeta = 2 \) yields exactly the same curves, which is unsurprising given that the divergences at each boundary share the same structure.

Moreover, in Fig.\ref{figure1}, plots are also shown for Dirichlet (top left) and Neumann (bottom left) boundary conditions at \( \zeta = 1 \) (\( x = 0 \)), with different values of \( \sigma \), including the case \( \sigma \rightarrow 0 \) represented by the solid black line. The latter reveals the position-dependent flight-time divergent behavior of the VD for fixed boundaries when \( x = 0 \). For this choice of position, the values of \( \theta \) corresponding to this type of divergence are the odd integers. In contrast, the divergences occurring at even values of \( \theta \) correspond to position-independent flight-time divergences arising from the first term on the r.h.s. of Eq. \eqref{VDR2}. It is worth noting that, for other values of \( x \), the position-dependent flight-time divergences will appear at different values of \( \theta \) in the plot. Additionally, it is evident from the plots that, for nonzero values of \( \sigma \), the divergences are smoothed out, as indicated by the corresponding curves.

In Fig.\ref{figure2}, the plots for the VD as a function of \( \theta \) are shown for the DN (top) and ND (bottom) cases. The discussion is similar to the situation illustrated in Fig.\ref{figure1}. Specifically, the plots on the right demonstrate that the VD is finite at the boundaries, while the plots on the left show the smoothing of the round-trip divergences, assuming \( \zeta = 1.5 \) (\( x = 0.5\bar{b} \)). If we assume \( \zeta = 1 \), as in the previous figure, the plots would exhibit the same behavior for both the DN and ND cases, since this value (\( x = 0 \)) corresponds to the midpoint between the boundaries, where the curves are symmetric.

We now turn to discuss the behavior of the VD in Eq. \eqref{VDRFluc} in some limiting cases. For instance, in the limit \( \sigma \rightarrow 0 \), we recover the fixed-boundaries case by using the asymptotic form for large arguments of the generalized hypergeometric function, i.e.,
\begin{eqnarray}
z\;_{2}F_{2}\left[1,1;\frac{3}{2},2;-z\right] \simeq \frac{1}{2}\ln(z) - \frac{1}{2}\psi(1/2),
\label{sigma0}
\end{eqnarray}
where \( \psi(y) \) is the digamma function, which is negligible compared to the logarithmic term. By substituting the above expression into Eq. \eqref{VDRFluc}, we can show that it reduces to Eq. \eqref{VDR2}. However, if we first evaluate the VD in Eq. \eqref{VDRFluc} at one of the round-trip divergences \( \theta = 2n + \zeta \) for a specific term \( n = N \) in the sum, and then apply the asymptotic limit above, we find that the VD diverges according to
\begin{eqnarray}
\overline{\langle(\Delta v)^2\rangle}_{\text{R}} \simeq -\frac{g^2}{4\pi m^2}\gamma_N^{(\text{i})}\ln\left[\frac{2(2N+1)^2\sigma^2}{(\zeta+2N)^2}\right].
\label{sigma_asymp}
\end{eqnarray}
In the case of a single-boundary (\( N = 0 \)) and adopting the Dirichlet condition (\( \gamma_0^{(\text{D})} = -1 \)), this has also been noted in Ref. \cite{de2014quantum}. On the other hand, if we first evaluate the VD in Eq. \eqref{VDRFluc} at one of the round-trip divergences \( \theta = 2n \) for a specific term \( n = N \) in the sum, it will diverge as
\begin{eqnarray}
\overline{\langle(\Delta v)^2\rangle}_{\text{R}} \simeq \frac{g^2}{4\pi m^2}\delta_N^{(\text{i})}\ln(2\sigma^2).
\label{sigma_asymp2}
\end{eqnarray}
The expression above, of course, does not apply to the single-boundary case. Both scenarios described in Eqs. \eqref{sigma_asymp} and \eqref{sigma_asymp2} are illustrated in the plots of Figs.\ref{figure1} and \ref{figure2} by the solid black lines, which clearly reveal these divergences.

Let us now turn to the calculation of the PD. To do so, we can more straightforwardly use Eq. \eqref{WFCompF}, along with its asymptotic form \eqref{logdiv}. Then, from Eq. \eqref{D_pos}, after performing the required differentiations and integrations, we obtain
\begin{eqnarray}
\langle(\Delta x)^2\rangle_{\text{R}}=\frac{g^2b^2}{2\pi m^2}\left[\sideset{}{'}\sum_{n=-\infty}^{\infty}\delta_n^{(\text{i})}E_n\left(\theta, 2n\right) -\sum_{n=-\infty}^{\infty}\gamma_n^{(\text{i})}E_n\left(\theta, \zeta + 2n\right)\right],
\label{RPD}
\end{eqnarray}
where
\begin{eqnarray}
E_n(\theta, r) = (\theta^2 - r^2)\ln\left[\frac{(\theta^2 - r^2)}{r^2}\right]^2 - 2\theta^2,
\label{Efun}
\end{eqnarray}
with \( \theta \) defined in Eq. \eqref{changeV}. We observe that the PD does not exhibit flight-time divergences, as the expression above remains finite in the limit \( \theta \rightarrow r \). This is consistent with the single-boundary case, which also lacks this type of divergence \cite{de2014quantum}. In Ref. \cite{Ferreira:2023uxs}, an interesting observation is made: in (3+1) dimensions, where two perfectly reflecting parallel planes impose Dirichlet, Neumann, and mixed boundary conditions on the scalar field, the component of the PD perpendicular to the planes exhibits flight-time divergences, while its parallel components remain finite.
\begin{figure}[h]
\includegraphics[scale=0.52]{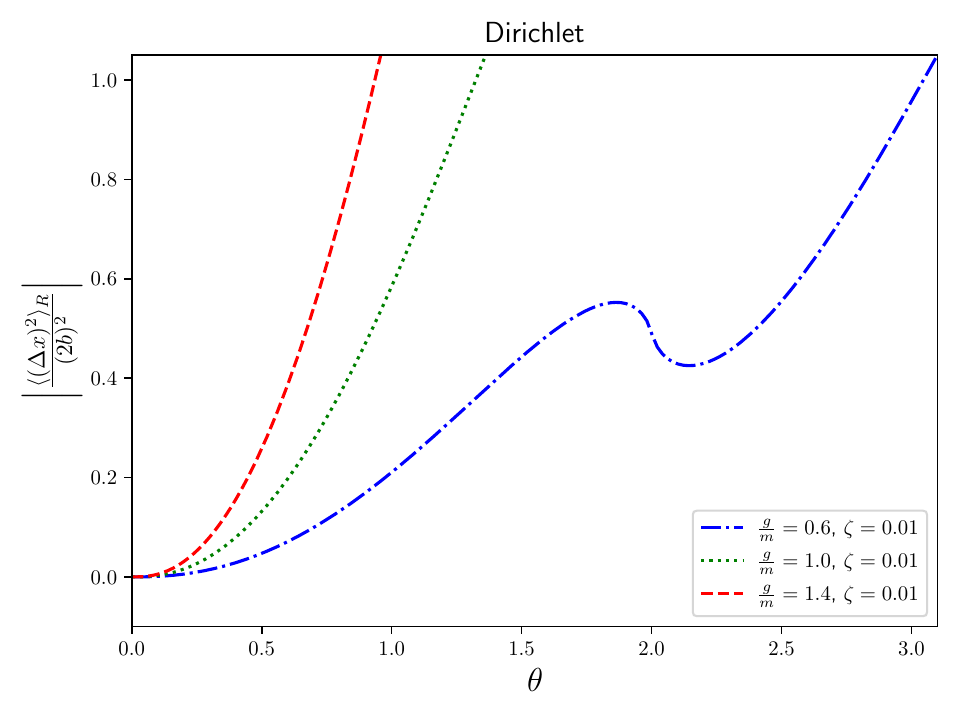}
\includegraphics[scale=0.52]{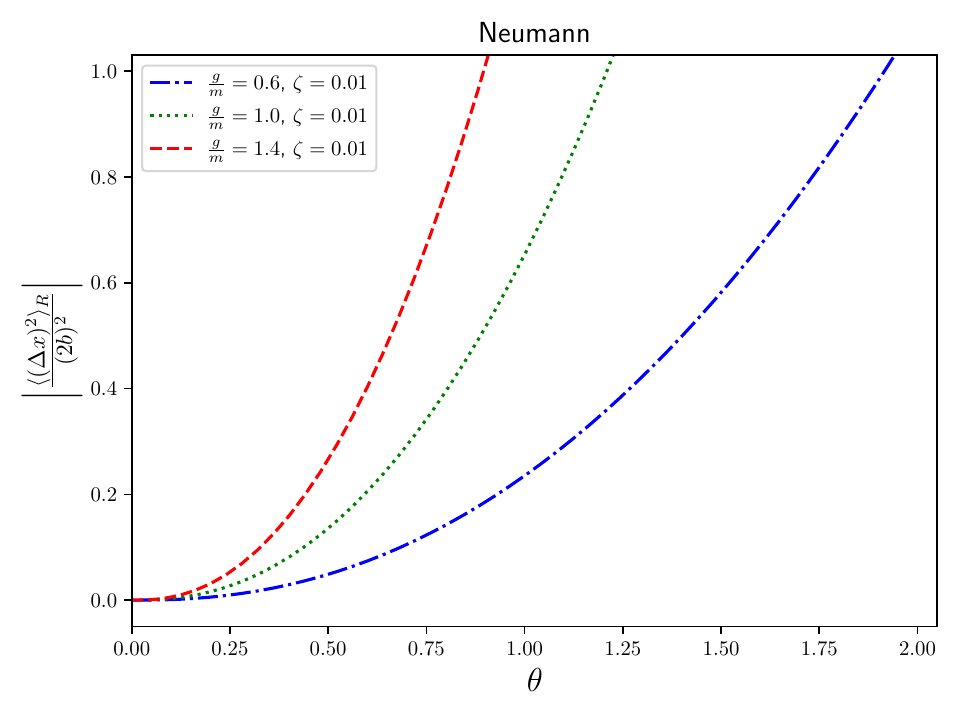}
\caption{Plots of the relative PD \eqref{SDC_DNM} as a function of \( \theta \), for Dirichlet and Neumann boundary conditions.}
\label{figure3}
\end{figure}

We still need to verify under what circumstances the SDC assumption holds. Thus, in order to justify the validity of this assumption, let us consider expanding the scalar field around the initial position $x_0$ of the particle:
\begin{equation}
\phi(x, t) = \phi(x_0, t) + (x - x_0)\partial_x \phi(x_0, t) + \frac{1}{2}(x - x_0)^2\partial_x^2 \phi(x_0, t) + \dots\,.
\label{justify}
\end{equation}
The expansion above, when inserted into Eq.~\eqref{vel}, the leading-order contribution to the velocity comes from the first spatial derivative, which we shall retain. The next-order term introduces a correction proportional to \(\langle (\Delta x)^2 \rangle\), after taking the VEV from quantization. Thus, neglecting this term is valid provided that \(\langle (\Delta x)^2 \rangle \ll \mathcal{L}^2\), where \(\mathcal{L}\) is the characteristic length scale of the system. 

Given that the length scale of the system considered above is the distance between the boundaries, \( 2b \), the SDC can be interpreted as requiring that the relative PD be much less than one, i.e.,
\begin{eqnarray}
\Bigg|\frac{\langle(\Delta x)^2\rangle_{\text{R}}}{(2b)^2}\Bigg| \ll 1.
\label{SDC_DNM}
\end{eqnarray}

In Fig.\ref{figure3}, we present the plots for the absolute values of the relative PD for Dirichlet and Neumann boundary conditions, with \( \zeta = 0.01 \). We observe that, in both cases, as the values of \( \frac{g}{m} \) decrease, the condition in Eq. \eqref{SDC_DNM} permits a wider range for \( \theta \). This is consistent with the investigations conducted in Refs. \cite{de2014quantum, Ferreira:2023uxs, JBFerreira:2023zwg}, where it is also shown that smaller values of \( g \) improve the validity of the SDC. Note that the same conclusion applies to the mixed boundary condition case. 

In Fig.\ref{figure3}, we also observe a monotonic increase of the position dispersion with time for most of the curves, which is consistent with a quantum version of random walk behavior. This reflects the accumulation of vacuum fluctuations over time. Interestingly, one of the plots (blue curve for Dirichlet) shows a temporary decrease in the position dispersion. We interpret this as a possible quantum interference effect arising from the nontrivial boundary conditions, where reflected modes can transiently cancel out the accumulated fluctuations. Such behavior may be related to revival-like phenomena arising from the confinement geometry, where interference among discrete field modes can temporarily reduce the quantum uncertainty in position, in a manner analogous to wave function revivals in quantum systems. A more detailed understanding of this phenomenon remains an open question and is left for future work. Note that this feature can also appear for other values of $\zeta$.

\subsection{Quasi-periodic condition}
%
We now wish to consider the Wightman function for the quasi-periodic condition in the form of Eq. \eqref{WFQuasi}. After performing the required integrations and derivatives in Eq. \eqref{D_vel} and taking the coincidence limit \( w' \rightarrow w \), we obtain
\begin{figure}[h]
\includegraphics[scale=0.3]{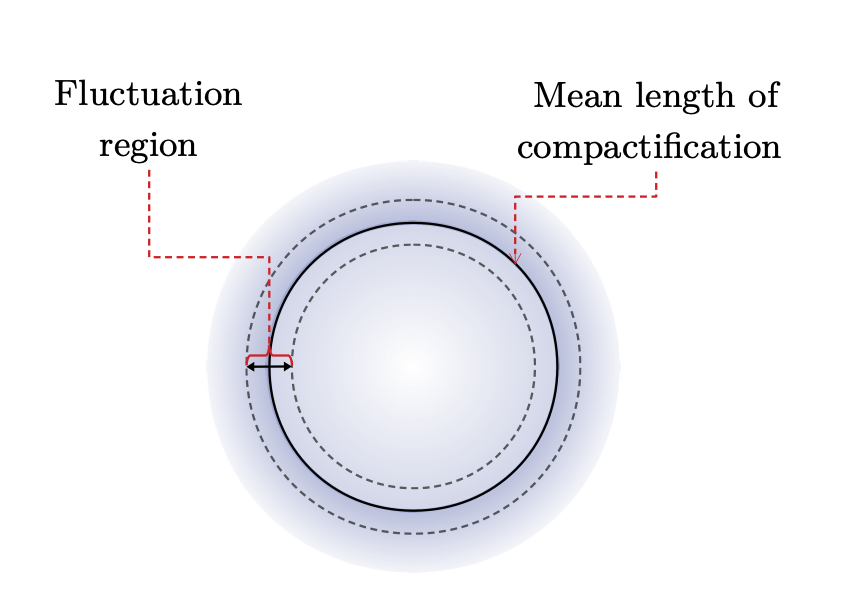}
\caption{Illustrative view of the fluctuating compactification length about their fixed mean length \( \bar{L} \).}
\label{figure01}
\end{figure}
\begin{eqnarray}
\langle(\Delta v)^2\rangle_{\text{R}}=\frac{g^2}{2\pi m^2}\sideset{}{'}\sum_{n=-\infty}^{\infty} e^{-2\pi in\beta} \int_{-\infty}^{\infty}d\kappa\frac{\kappa^2\left[1-\cos(|\kappa|\tau)\right]}{|\kappa|^3}e^{i\kappa nL},
\label{VDRquasi1}
\end{eqnarray}
where we have used $\omega_{\kappa}=|\kappa|$. Thus, the integral above can be calculated by using Eq. \eqref{int1}. This provides
\begin{eqnarray}
\langle(\Delta v)^2\rangle_{\text{R}}=-\frac{g^2}{2\pi m^2}\sum_{n=1}^{\infty} \cos(2\pi n\beta)\ln\left(\frac{n^2}{n^2 - \tau_L^2}\right)^2,
\label{VDRquasi2}
\end{eqnarray}
with $\tau_L=\frac{\tau}{L}$. We observe that the expression above diverges for \( \tau = nL \). Similar to the case discussed in the previous subsection, where the divergences are associated with the time a light signal takes to complete multiple round trips between the boundaries, here we can interpret the divergences as arising from the time a light signal takes to travel multiple times around a circle of length \( L \), i.e., the compactification length. This type of divergence also occurs in (3+1)-dimensional spacetime, as demonstrated in Ref. \cite{Ferreira:2023uxs}.
\begin{figure}[h]
\includegraphics[scale=0.55]{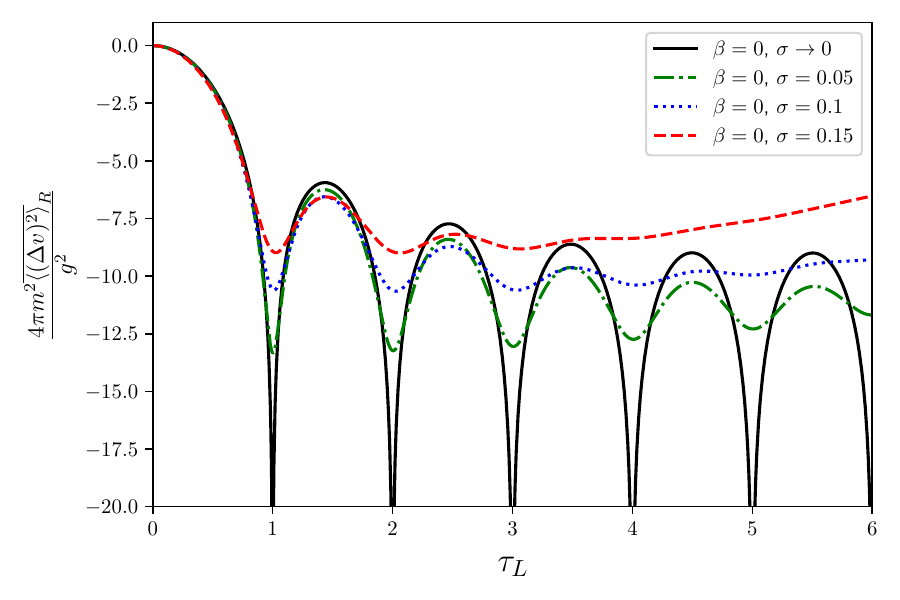}
\includegraphics[scale=0.55]{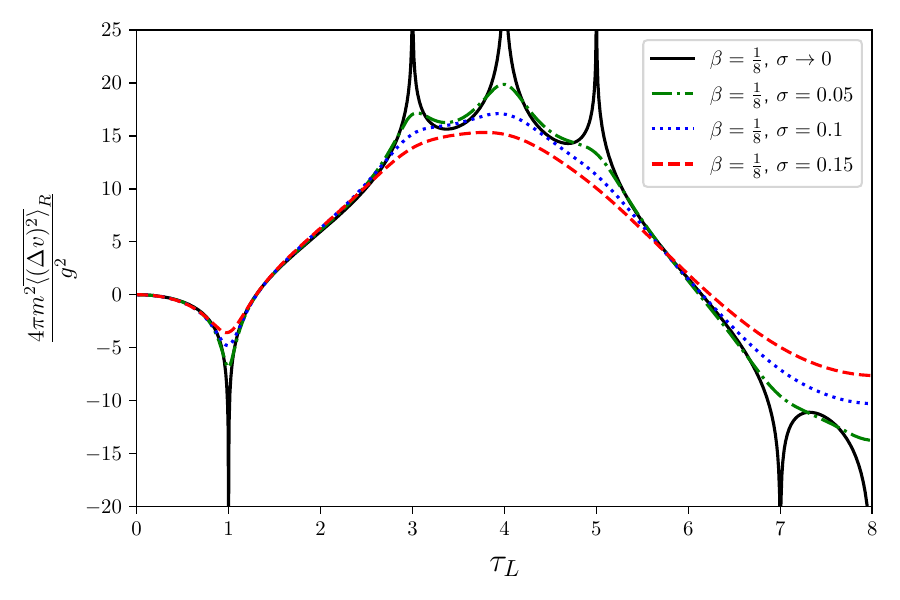}
\includegraphics[scale=0.55]{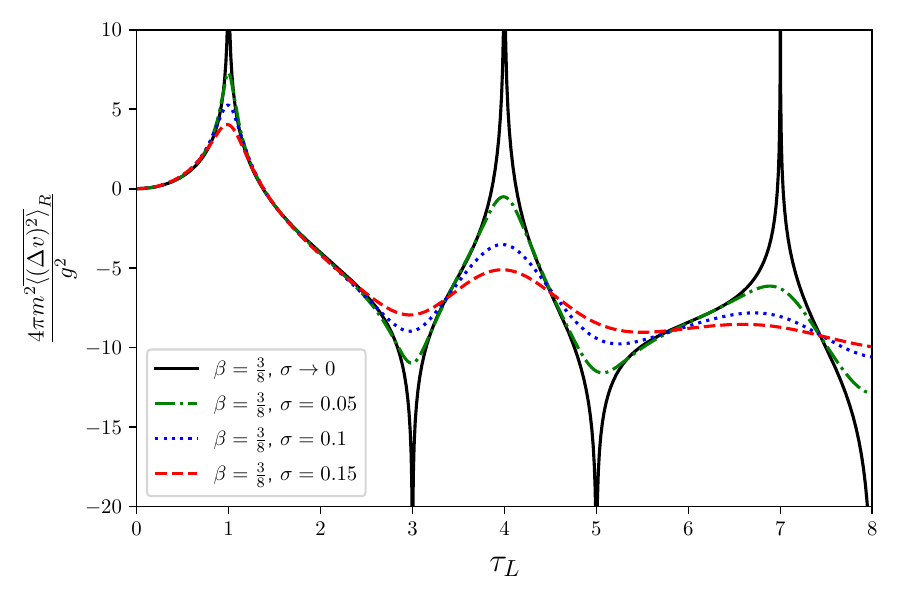}
\includegraphics[scale=0.55]{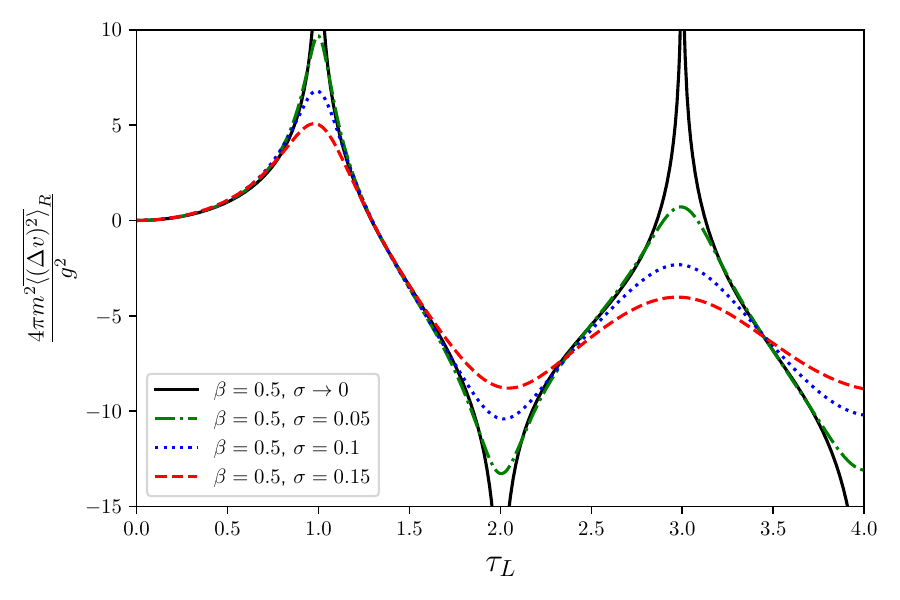}
\caption{Plots of the VD \eqref{VDRFlucquasi} in terms of $\tau_L$,  considering periodic ($\beta=0$), anti-periodic ($\beta=0.5$) and other interpolated values of $\beta$. It is clear in each case that for finite values of $\sigma$ the flight time divergences are smoothed out.}
\label{figure4}
\end{figure}

To smooth out the divergences in Eq. \eqref{VDRquasi2}, we can treat the compactification length \( L \) as a random variable that fluctuates according to a Gaussian distribution about its mean value \( \bar{L} \), similar to Eq. \eqref{b_fluctuation} (see Fig.\ref{figure01}). By using Eqs. \eqref{PDG} and \eqref{MV}, we obtain
\begin{eqnarray}
\overline{\langle(\Delta v)^2\rangle}_{\text{R}}=-\frac{g^2}{2\pi m^2\sigma^2}\sum_{n=1}^{\infty}\cos(2\pi n\beta)Q_n(\tau_L), 
\label{VDRFlucquasi}
\end{eqnarray}
where we have defined the function
\begin{eqnarray}
Q_n(\tau_L)=&-&\left(1 + \frac{\tau_L}{n}\right)^2\;_{2}F_{2}\left[1,1;\frac{3}{2},2;-\frac{\left(1 + \frac{\tau_L}{n}\right)^2}{2\sigma^2}\right]\nonumber\\
&-&\left(1 - \frac{\tau_L}{n}\right)^2\;_{2}F_{2}\left[1,1;\frac{3}{2},2;-\frac{\left(1 - \frac{\tau_L}{n}\right)^2}{2\sigma^2}\right]\nonumber\\
&+&2\;_{2}F_{2}\left[1,1;\frac{3}{2},2;-\frac{1}{2\sigma^2}\right].
\label{p1}
\end{eqnarray}

The expression in Eq. \eqref{VDRFlucquasi} is now finite at \( \tau = nL \). This is demonstrated in the plots of Fig.\ref{figure4} for different values of \( \sigma \) and the quasi-periodicity parameter \( \beta \). These plots also illustrate the transition from the well-known periodic condition (\( \beta = 0 \)) to the anti-periodic condition (\( \beta = 1/2 \)) by considering intermediate values such as \( \beta = 1/8 \) and \( \beta = 3/8 \). We observe that the periodic case closely resembles the Dirichlet boundary condition case shown in Fig.\ref{figure1}, while the anti-periodic case is very similar to the Neumann boundary condition case. The value $\beta=1/8$ exhibits a transitional curve shape that bears some resemblance to the mixed ND condition case depicted in Fig.\ref{figure2}. However, it is important to note that the Dirichlet, Neumann, and mixed boundary condition cases analyzed in the previous subsection are fundamentally distinct from the quasi-periodic condition case. The quasi-periodic condition results in the topology of the spacetime becoming \( \mathbb{R}^1 \times S^1 \), whereas the former cases lead to an anisotropic spacetime \cite{Ferreira:2023uxs}.

Moreover, using the asymptotic form of the generalized hypergeometric function in Eq. \eqref{sigma0}, we can demonstrate that the VD in Eq. \eqref{VDRFlucquasi} reduces to the one in Eq. \eqref{VDRquasi2} for a fixed length $L$. Additionally, we can verify that when evaluated at $\tau=nL$ and using \eqref{sigma0},
Eq. \eqref{VDRFlucquasi} diverges in a manner similar to Eq. \eqref{sigma_asymp2} as $\sigma\rightarrow 0$. 
\begin{figure}[h]
\includegraphics[scale=0.52]{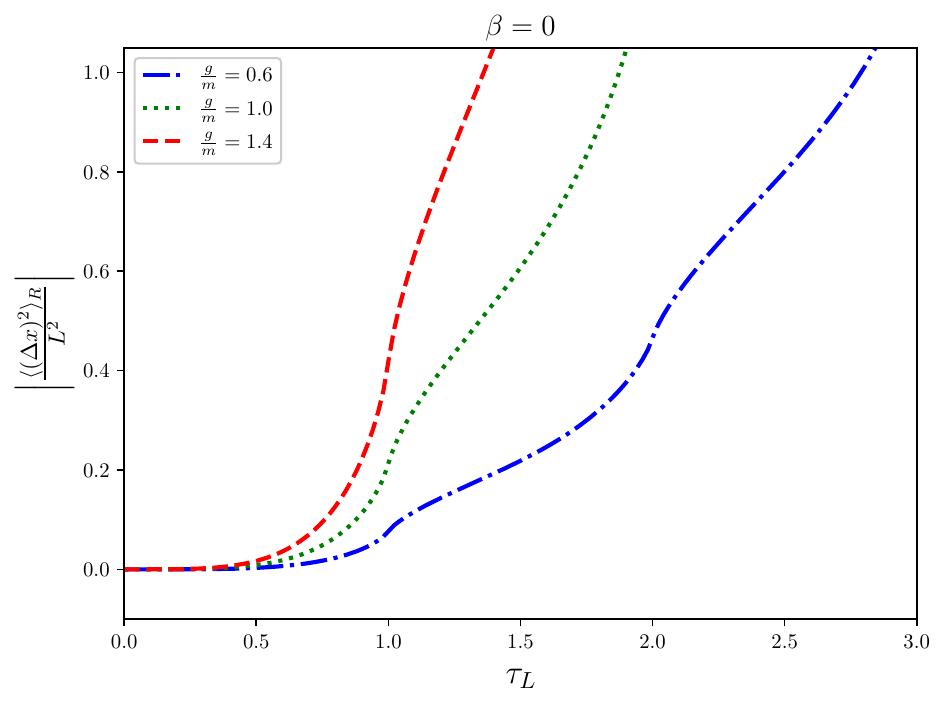}
\includegraphics[scale=0.52]{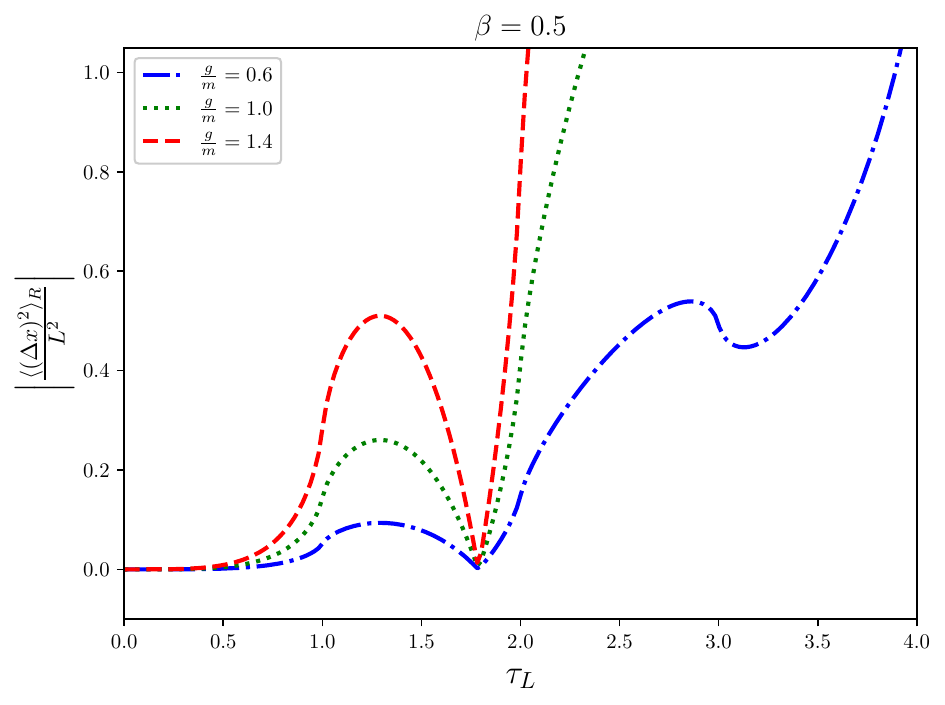}
\caption{Plots of the relative PD \eqref{SDC_quasi} as a function of \( \theta \), for the periodic and anti-periodic condition cases.}
\label{figure5}
\end{figure}

The PD of the particle induced by the quasi-periodic condition can be calculated using the Wightman function \eqref{WFQuasi2}, along with its corresponding asymptotic form \eqref{logdiv} in the massless scalar field case. Consequently, from Eq. \eqref{D_pos}, after performing the required differentiations and integrations, we obtain
\begin{eqnarray}
\langle(\Delta x)^2\rangle_{\text{R}}=\frac{g^2L^2}{4\pi m^2}\sum_{n=1}^{\infty}\cos(2\pi n\beta)E_n\left(\tau_L, n\right),
\label{RPDquasi}
\end{eqnarray}
where the function \( E_n(\tau_L, n) \) has the form of Eq. \eqref{Efun} with \( \theta \) replaced by \( \tau_L \). The expression above is also free from flight-time divergences at \( \tau = nL \). In Ref. \cite{Ferreira:2023uxs}, it has also been shown that, in the (3+1)-dimensional case, the component of the PD along the compactified dimension exhibits flight-time divergences at \( \tau = nL \), while the components along the uncompactified dimensions remain regular.

We can now discuss the SDC assumption based on the expansion in Eq.~\eqref{justify}. In this context, the relevant length scale of the system is the compactification length \( L \). Consequently, the relative PD must be expressed as
\begin{eqnarray}
\Bigg|\frac{\langle(\Delta x)^2\rangle_{\text{R}}}{L^2}\Bigg| \ll 1.
\label{SDC_quasi}
\end{eqnarray}

In Fig.\ref{figure5}, we present the plots for the absolute values of the relative PD for the periodic (\( \beta = 0 \)) and anti-periodic (\( \beta = 0.5 \)) condition cases. As expected, we observe that, in both cases, decreasing the values of \( \frac{g}{m} \) allows for a wider range of \( \tau_L \), consistent with the condition in Eq. \eqref{SDC_quasi}. We observe that, unlike the blue curve (Dirichlet case) in Fig.\ref{figure3}, the temporary decrease in the position dispersion is more pronounced for the antiperiodic condition. In contrast, the periodic case exhibits mildly oscillatory behavior without an actual decrease in the dispersion. Such oscillations are still compatible with random walk-like behavior and are expected in confined quantum systems due to interference among discrete field modes. It is also worth noting that a temporary decrease in the position dispersion can be observed for other intermediate values of $\beta$, not only in the strictly antiperiodic case. 

\section{Conclusions and final remarks}
\label{sec4}

We have explored the quantum Brownian motion of a point charge driven by quantum fluctuations of a real massless scalar field under Dirichlet, Neumann, mixed boundary conditions, and a quasi-periodic condition. By introducing fluctuations in the positions of two point-like boundaries and the compactification length, we generalized and extended previous findings in the literature. Our results demonstrate that the typical divergences linked to fixed boundaries and compactification are effectively smoothed out when a wave function is associated with the characteristic length scale of each system.

The results have shown that the VD of the point charge can assume negative values, a phenomenon known as subvacuum effects, which arises from the reduction in quantum uncertainty due to the presence of boundaries. Furthermore, the flight-time divergences that appear in scenarios with fixed boundaries are regularized when the boundaries are treated as quantum objects with an associated wave function. This has been achieved by considering a Gaussian distribution for the position of the boundaries and the compactification length, resulting in finite expressions for the velocity and PDs of the particle.

For Dirichlet, Neumann, and mixed boundary conditions, we have demonstrated that flight-time divergences and divergences at the boundary positions are smoothed out when the boundaries fluctuate. In the case of the quasi-periodic condition, the divergences associated with the time it takes for a light signal to travel multiple times the compactification length are also regularized. These results have been corroborated by graphical analyses showing the smoothed behavior of the VD as a function of time and position in the case of Dirichlet, Neumann, and mixed boundary conditions, and as a function of time and quasiperiodicity parameter $\beta$ in the case of a quasi-periodic condition. 

Additionally, we have verified that the small displacement condition assumption holds for a wide range of parameters, especially when the ratio between the charge and mass of the particle is small. This ensures that the relative position dispersion is much less than one, maintaining the consistency of the model.

In summary, this work provides a more comprehensive understanding of quantum Brownian motion in systems with fluctuating boundaries and compactification, highlighting the importance of treating boundaries as quantum entities to regularize divergences and explore non-trivial quantum effects. These results have implications for the study of quantum fluctuations in confined geometries. In the final remarks of Ref. \cite{DeLorenci:2016jhd}, the authors argue that the regularization method for the typical divergences using a Gaussian distribution is not effective when the system studied in \cite{de2014quantum}, where a single point-like boundary with Dirichlet boundary condition was considered, is generalized to higher dimensions. Despite this, it is still worth considering as a future investigation whether this limitation also holds for the generalization of the systems studied here to higher dimensions. Additionally, it may be worthwhile to explore this issue for other configurations, as the effectiveness of such regularization methods could vary depending on the specific boundary conditions and geometric setups.
{\acknowledgments}

The authors are grateful to \'{E}werton J.  B. Ferreira for helpful discussions. E.M.B.G would like to thank the Brazilian agency Coordination for the Improvement of Higher Education Personnel (CAPES) for financial support. H.M is partially supported by the National Council for Scientific and Technological Development (CNPq) under grant No 308049/2023-3.


\end{document}